\documentclass[titlepage,11pt]{article}
\usepackage{etoolbox, mathtools}

\makeatletter
\patchcmd{\@makecaption}
  {\parbox}
  {\advance\@tempdima-\fontdimen2} 
  {}{}
\makeatother  

\usepackage[caption = false]{subfig} 
\usepackage{amsmath}

\usepackage{anyfontsize}
\usepackage[a4paper]{geometry}
\usepackage{graphicx}
\usepackage[textwidth=8em,textsize=small]{todonotes}
\usepackage{amssymb,euscript,amsfonts,mathrsfs,amscd}
\usepackage{natbib}
\RequirePackage{hyperref}
\usepackage{multirow}

\definecolor{lavender}{RGB}{180, 52, 235}

\title{A Seamless Phase I/II Platform Design with a Time-To-Event Efficacy Endpoint for Potential COVID-19 Therapies}
\author{Thomas Jaki$^{1,2}$ and Helen Barnett$^{1}$ and  Andrew Titman$^{1}$ and  Pavel Mozgunov$^{1}$\\\\$^1$ Department of Mathematics and Statistics, Lancaster University,
Lancaster,
UK.\\
$^2$ MRC Biostatistics Unit, University of Cambridge, Cambridge, UK.}

\begin{document}
\maketitle

\begin{abstract}
In the search for effective treatments for COVID-19, initial emphasis has been on re-purposed treatments. To maximise the chances of finding successful treatments, novel treatments that have been developed for this disease in particular, are needed. In this manuscript we describe and evaluate the statistical design of the AGILE platform, an adaptive randomized seamless Phase I/II trial platform that seeks to quickly establish a safe range of doses and investigates treatments for potential efficacy using a Bayesian sequential trial design. Both single agent and combination treatments are considered. We find that the design can identify potential treatments that are safe and efficacious reliably with small to moderate sample sizes.\\\\

\end{abstract}

\clearpage
\section{Introduction}
The emergence of COVID-19 and the ensuing pandemic has led to a widespread, frantic, search for treatments. Despite large uncertainty about the underlying pathogen and the natural history of disease, trials must start rapidly to identify
treatments to save lives, but also so that effective treatments can be used in the response to the outbreak. A consequence of this is that trials in COVID-19 in the first few weeks and months of the outbreak have focused on re-purposed treatments~\citep{cao2020trial,wang2020remdesivir,HorbyHCQ}.\\

While recently, some success with using re-purposed treatments has been achieved \citep{beigel2020remdesivir,horbydexamethasone}, it is crucial that development of treatments specifically developed for COVID-19 is also undertaken in order to maximise the chances finding therapies to successfully treat patients. The crucial difference of trials investigating novel therapies (in contrast to re-purposed treatments) is that the range of safe and likely effective doses is unknown. Therefore, an efficient dose-finding design identifying safe and active doses to be studied in larger trials is essential. While there exist a number of dose-finding design for early-phase dose-finding trials evaluating toxicity and efficacy simultaneously~\citep[see e.g.][and references therein]{wages2015,mozgunov2018information}, many of them consider a binary efficacy endpoint~\citep[with few recent extension to continious endpoints, see e.g.][]{hirakawa2012adaptive,mozgunov2019flexible}. Time-to-event endpoints with censoring at 28-days have, however, previously been used as a clinically meaningful measure in a number of COVID-19 trials~\citep{cao2020trial,wang2020remdesivir,beigel2020remdesivir} and the argument has been made that they should be considered in COVID-19 trials~\citep{dodd2020}.\\

While the majority of Phase I dose-finding trials are non-randomised, it is agreed that in later phases, the gold standard for evaluating novel treatments are well conducted randomized controlled clinical trials. At the same time, in the light of the uncertainty about the adverse events caused by COVID-19, it is essential to conduct a randomised dose-finding trial to ensure that the risk of adverse events is correctly attributed to the drug under study rather than to the disease itself. Moreover, it has been argued that adaptive designs \citep[e.g.][]{pallmann2018adaptive} are particularly suitable during a pandemic, also in the light of the uncertainty about a novel disease~\citep{stallard2020efficient}. Therefore, a randomised adaptive dose-finding design evaluating both toxicity and time-to-event efficacy would allow to answer the research questions of interest in novel therapies for treating COVID-19.\\

It is also recognised that there is a number of novel therapies that have a potential to be efficient in fighting the COVID-19. Therefore, it is crucial to have a structure in place that would allow rapid enrolment of novel therapies to ensure rapid decision-making, and, importantly, would allow for effecient use of information between the studies, i.e. utilising the data from the control treatment across different compounds. This can be achieved via a platform trial~\citep{meyer2020evolution}. \\

In this paper, we describe and evaluate the design developed and implemented for the AGILE platform \citep{griffiths2020agile}, an adaptive randomized seamless Phase~I/II dose-finding trial platform that seeks to quickly establish a safe range of doses and investigates treatments for potential efficacy using a Bayesian sequential trial design. The proposed design uses statistical models to improve a decision making, and further efficiency is gained by sharing control group patients between concurrent controls across different candidates in the platform. We also extend the design for the trials studying dual-agent combination of treatments. \\

The rest of manuscript proceeds as follows.  Section \ref{sec:singlemethods} describes the platform for single treatments while its performance is evaluated in simulations in Section \ref{sec:singleresults}. The design for dual-agent combinations is proposed in Section \ref{sec:combomethods} and subsequently evaluated in Section \ref{sec:comboresults}. We conclude with a discussion (Section \ref{sec:discussion}).

\section{Single agent design \label{sec:singlemethods}}

\subsection{Setting \label{sec:single-agent-setting}}

Consider a randomized controlled dose-escalation clinical trial in which $m$ increasing doses $d_1 < d_2 < \ldots < d_m$ of a single experimental treatment are studied. Let ${d}_0=0$ be a dose of zero of the treatment, which is subsequently referred to as the \textit{control arm} (or, simply, control). The inclusion of the control arm is motivated by the still emerging nature of the symptoms associated with COVID-19 and the desire to avoid labelling potential treatments as unsafe due to misclassifying non-treatment related symptoms. A binary outcome of a random variable $Y$, $y=0$ is observed if no dose-limiting event (DLE) is observed within $t_{safe}>0$ days after randomisation, and $y=1$, otherwise. Let $p_j$ be the probability for a patient to experience a DLE if given dose ${d}_j$. It is assumed that the risk of a DLE is a non-decreasing function of dose, $p_0 \leq p_1 \leq \ldots \leq p_m$ and prior information for the DLE probability of the control arm, $p_0$, is available.\\

As it is expected that the control arm is associated with a non-negative (unknown) risk of DLE (or symptoms of the disease that can not be distinguished from DLEs), the primary goal of the dose-escalation is formulated in terms of the additional risk of a dose limiting event (ADLE) defined in terms of the expected difference in DLE risk between the doses of the agent and the control. Specifically, we therefore seek to identify the dose that corresponds to an ADLE risk of $\gamma=0.20$ which equates to finding the dose $d_{j^\star}$ such that
$$j^\star=\arg \min_{j=0,\ldots,m} | \left(p_j-p_0 \right) -\gamma|.$$

\subsection{Bayesian Dose-Escalation Model}
\label{sec:single}
The following randomized Bayesian dose-escalation design that builds on the proposal by~\cite{mozgunov2019randomised} is used. Assume that the DLE probability has the functional form
\begin{equation}
\psi(\tilde{d}_j,\theta_1,\theta_2) = \frac{\exp(\theta_1 + \theta_2\tilde{d}_j)}{1+\exp(\theta_1 + \theta_2\tilde{d}_j)}
\label{eq:toxicity}
\end{equation}
where $\theta_1$ and $\theta_2$ are unknown parameters, and $\tilde{d}_j$ is a standardized dose level (also referred to as a skeleton) corresponding to dose $j$, which are constucted given the information about the prior DLE toxicities (details are given below). This model choice was found to result in good statistical properties in terms of the target dose identification in a randomized dose-finding trial~\citep{mozgunov2019randomised}. We require that the standardized dose level corresponding to control is equal to $\tilde{d}_0=0$. This will guarantee that a sequential update of the slope parameter $\theta_1$ will not contribute to the DLE probability estimation on the control arm yet all data are used for its estimation~\citep{mozgunov2019randomised}.\\

Denote the prior distribution of the vector $\theta=(\theta_1,\theta_2)$ by $f_0(.)$. To construct the standardized levels, $\tilde{d}_j$, we represent them in terms of prior estimates of the DLE probabilities $\hat{p}_j^{(0)}$ associated with doses ${d}_j$ $j=0,\ldots,m$
\begin{equation}\tilde{d}_j= \frac{{\rm logit} (\hat{p}_j^{(0)}) -\hat{\theta}_1^{(0)} }{\hat{\theta}_2^{(0)}}
\label{eq:skeleton}
\end{equation}
where $\hat{\theta}_1^{(0)}$, $\hat{\theta}_2^{(0)}$ are prior point estimates of the model parameters, and ${\rm logit}(x)=\log \frac{x}{1-x}$ is the logit transformation of $x$. To satisfy $\tilde{d}_0=0$, the prior needs to be chosen such that ${\rm logit}(\hat{p}_0^{(0)})= \hat{\theta}_1^{(0)}$.\\

Assume that $n$ patients have already been assigned to doses $\tilde{d}(1), \ldots, \tilde{d}(n)$ and binary responses $\mathbb{Y}_n=[y_1, \ldots, y_n]^{\rm T}$ were observed, respectively. The models updates the posterior distribution of $\theta$ using Bayes' Theorem
\begin{equation}
f_n(\theta)= \frac{f_{n-1}(\theta)\phi(\tilde{d}(n),y_n,\theta)}{\int_{\mathbb{R}^h} f_{n-1}(u)\phi(\tilde{d}(n),y_n,u) {\rm d} u}=\frac{f_0(\theta) \prod_{i=1}^{n}\phi(\tilde{d}(i),y_i,\theta)}{\int_{\mathbb{R}^h}  f_0(u) \prod_{i=1}^{n}\phi(\tilde{d}(i),y_i,u) {\rm d} u}
\label{posterior}
\end{equation}
where
$\phi(d(n),y_n,\theta)=\psi(d(n),\theta)^{y_n}(1-\psi(d(n),\theta))^{1-y_n}.$
This posterior distribution is then used to make the escalation/de-escalation decision~\citep{beat2008}. Specifically, the first set of safe doses is defined as the doses $d_j$ for which 
\begin{equation}
\mathbb{P} \left( p_j-p_0  \geq \gamma+2\delta \right) < c_{\rm overdose}
\label{eq:safety}
\end{equation}
where $\gamma$ is the target ADLE risk, $\delta$ is the width of the interval of DLE risk which we consider acceptable, $c_{\rm overdose}$ is the threshold controlling overdosing, and the probability is found with respect to the updated posterior distribution. Amongst the safe doses, the dose which maximises
\begin{equation}
\mathbb{P} \left( p_j-p_0 \in [\gamma-\delta,\gamma+\delta] \right)
\label{eq:criterion}
\end{equation}
is selected as the target dose.

\subsection{Efficacy Design}

\subsubsection{Baeysian Efficacy Model}

For the efficacy endpoint, a Cox proportional hazards model is assumed where the hazard of recovery at time $t$ is given by $h(t ; z) = h_0(t ; z)\psi^z$ where $z=1$ corresponds to a treatment and $z=0$ to control. Initially the cohort of patients who have graduated from the dose-escalation stage are followed up for a total of $t_{eff}$ days. Based on their outcomes, a decision is made to either stop for futility, stop for efficacy or recruit a further cohort of patients. To improve power, controls recruited from other candidate treatments or other doses of the same treatment are also used within the evaluation, but this is restricted to using only the most recent $n_c$ such controls to mitigate the risk of bias due to population drift. \\

A Bayesian criterion is adopted for the stopping rule at each stage. In line with Bayesian thinking, we set the stopping rules to be the same for each stage $k$. Specifically denoting all data up to stage $k$ on dose $j$ by $\mathcal{D}^{(j)}_k$ and for a given desirable treatment effect, $\psi^{*}$,
\begin{itemize}
\item evaluation is stopped for efficacy if $\pi^{(j)}_{E | k} \coloneqq \mathbb{P}(\psi=\psi^{*} \mid \mathcal{D}^{(j)}_k)  > u$, 
\item evaluation is stopped for futility if $\pi^{(j)}_{E | k} < l$, or
\item an additional cohort of patients is recruited, otherwise. 
\end{itemize}
At the final stage $k=K$, efficacy for dose $j$ is established if $\pi^{(j)}_{E | K} > u$.\\

A point prior of the form $\mathbb{P}(\psi=1) = 1 - \pi^{(j)}_E,~ \mathbb{P}(\psi=\psi^{*})=\pi^{(j)}_E$ is assumed for $\psi$. Here $\pi^{(j)}_E$ represents the degree of optimism or scepticism towards the likely efficacy of dose $d_j$. \\

An advantage of the point prior is that obtaining the posterior probability $\pi^{(j)}_{E | k}$ is computationally straightforward allowing comprehensive evaluations of the design via simulations. The posterior probability under this model is
\begin{equation}\pi^{(j)}_{E | k} = \frac{\pi^{(j)}_E L(\psi^{*} \mid \mathcal{D}^{(j)}_k)}{\pi^{(j)}_E L(\psi^{*} \mid \mathcal{D}^{(j)}_k) + (1-\pi^{(j)}_E) L(1 \mid \mathcal{D}^{(j)}_k)},\label{post_prob}\end{equation}
where $L( u \mid \mathcal{D}^{(j)}_k)$ is the Cox partial likelihood with respect to the data for dose $j$ up to the $k$th period evaluated at a hazard ratio of $u$. While the Cox partial likelihood is often not considered compatible with a Bayesian analysis since it does not use the full information of the data, Bayesian justifications of its use are available \citep{kalbfleisch, sinha}.

\subsubsection{Setting the boundaries}

To set the boundaries, $(l, u)$, a large number of trajectories of $L( 1 \mid \mathcal{D}^{(j)}_k)$ and  $L( \psi^{*} \mid \mathcal{D}^{(j)}_k)$ can be simulated under both the null and alternative hypothesis, where in all cases the simulation continues until the maximum period $K$. An assumption regarding the proportion of patients recovered by time $t_{eff}$ under the null is needed, in addition to the hazard ratio between treatments. In the absence of censoring due to drop-out, the results will be otherwise invariant to the precise survival distribution assumed since the Cox partial likelihood only uses the order of events.\\

The effect of varying $(l, u)$ can then be explored by converting the pairs of likelihoods into a posterior probability and imposing the boundary stopping rules. For any given set of boundaries, the type I error, power, expected number of patients under the null, expected number of patients under the alternative and probabilities of stopping for futility or efficacy at each stage, can be approximated. The boundaries can then be set to optimize some criterion, for instance sum of expected sample sizes under the null and alternative, subject to some constraints, for instance controlling type I error, keeping power above some level or limiting the chance of early stopping under the alternative. \\

The inclusion of historic controls will increase both the power and type I error of any procedure \citep[e.g.][]{schmidli2014robust}. To ensure type I error is controlled, the boundaries are set assuming the maximum, $n_c$, previous controls are available, with the consequence that the type I error will be lower for the first treatment evaluated in the platform. This will also mean the power will be lower for the first few evaluated treatments. However, given the dose finding design, it is anticipated that the first evaluations will be of less importance, as for safety reasons evaluations tend to start at sub-optimal low doses.

\subsection{Overall Design \label{design}}

Throughout the study, patients are allocated in cohorts of size $c=c_1+c_2$ where $c_1$ is the number of patients in the cohort assigned to an active dose and $c_2$ is the number of patients in cohort assigned to the control arm, $\tilde{d}_0$. Below is an outline of the overall procedure made up by both safety and efficacy evaluation:\\

\underline{Safety-evaluation}
\begin{enumerate}
\item The first cohort of $c_1+c_2$ patients is assigned to the first dose and to the control arm, respectively.
\item After $t_{safe}$ days, short-term DLE outcomes are collected and the posterior distribution of the parameters is updated using Equation~\ref{posterior}.
\item The set of safe doses is found using Equation~\ref{eq:safety}.
\begin{itemize}
\item If no doses are safe, the trial is stopped for safety;
\item if only the current dose is safe, the next cohort of $c_1+c_2$ patients is assigned to the current dose and to the control arm, respectively;
\item otherwise, the next cohort of $c_1+c_2$ patients is assigned to the {adjacent, safe} dose level for which the probability~(\ref{eq:criterion}) is maximized and control arm respectively.
\end{itemize}
\item Once efficacy information is available for two cohorts on a safe dose, that dose is graduated to the efficacy evaluation. 
\end{enumerate}

\underline{Efficacy-evaluation}
\begin{enumerate}
 \item If a dose $d_j$ is deemed safe, the efficacy outcome is observed up to day, $t_{eff}$. 
 \item The posterior probability, $\pi^{(j)}_{E|k}$, following equation \ref{post_prob} is then computed where $k$ corresponds to the number of times this dose has been evaluated for efficacy.
 \begin{itemize}
  \item If $\pi^{(j)}_{E|k}<l$ evaluation of dose $d_j$ is stopped for futility;
 \item if $\pi^{(j)}_{E|k}>u$ evaluation is stopped for efficacy and the corresponding candidate and dose $d_j$ recommended for further testing;
 \item otherwise if $k<K$, an additional cohort of $c_1 +c_2$ patients is recruited on the current dose and control arm, respectively.
 \end{itemize}
\end{enumerate}

\begin{figure}[!ht]
\begin{center}
\includegraphics[width=0.8\linewidth]{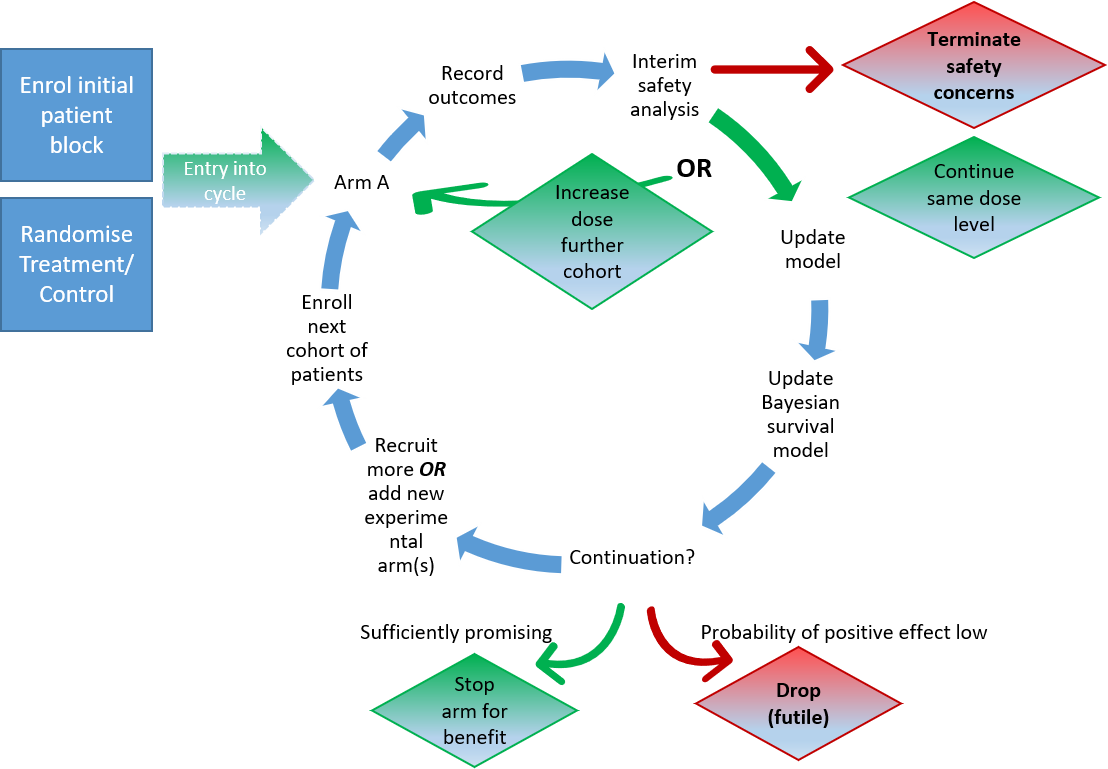}
\caption{Illustration of the AGILE platform design.\label{fig:AGILE}}
\end{center}
\end{figure}

The evaluation of a dose continues until the maximum number of patients $N$ on a dose has been reached unless it is stopped for efficacy, futility or safety before. Once all doses are stopped, the evaluation of this candidate stops. The overall design of the study is depicted in Figure \ref{fig:AGILE}. \\

\section{Evaluations of Proposed Design\label{sec:singleresults}}
\subsection{Setting}

We will now evaluate safety and efficacy across the study together in a simulation study and evaluate the impact of shared control data that are gradually accumulated over the course of the trial. \\

We consider the setting where there are three active doses ($m=3$) of a single agent and the control arm. As before, the DLE endpoint is binary and corresponds to either experiencing or not experiencing a DLE by time $t_{safe}=7$ and the efficacy endpoint is time-to-improvement defined as a 1 category improvement on the 10-point WHO scale \citep{marshall2020} over 28 days, $t_{eff}=28$. To generate the efficacy outcome a Weibull distribution is used. For the control group the rate and shape parameters were 0.085 and 0.797, respectively, resulting in a median recovery time of 14 days and a recovery rate of 70\% within 28 days. The rate parameter for the candidate dose have been adapted to match the scenarios described below. The binary safety and time-to-event efficacy responses are assumed to be highly correlated and are generated via a procedure described by~\cite{mozgunov2018benchmark} using a correlation coefficient $\rho=0.8$.\\

The maximum total intake per dose level is 72 patients assigned to each dose level and control, which equates to a maximum total sample size of 216. In line with the real study, cohort sizes of $c_1=4$ (assigned to active doses) and $c_2=2$ (assigned to the control) are used. Moreover, we also report sensitivity analysis to assess the impact of various cohort size.\\ 

The objective of the trial is to find all safe efficacious doses to be graduated into a larger Phase II or Phase III clinical trial. The target ADLE risk is $\gamma=0.20$, and the dose is considered safe if the ADLE risk is less than or equal to $0.30$ ($\delta=0.05$) and is efficacious if it corresponds to hazard ratio of at least 1.75. The stopping boundaries for efficacy have been found such that the type I error in each pairwise comparison is controlled at 10\% one-sided, while the power for each individual dose is 80\%. The resulting boundary values are $l=0.224$ and $u=0.839$ (see Section 3.3.2 for more details).

\subsection{Scenarios}\label{sec:scen_single}

As the trial aims to study novel compounds which have yet to be explored with respect to their mechanism of action in COVID-19 patients, it is crucial that the design has good operating characteristics under a variety of dose-DLE and dose-efficacy scenarios. Therefore, we consider 5 dose-efficacy scenarios ranging from no doses corresponding to a change in time-to-improvement within 28-days, to all doses resulting in a clinically significant reduction; and 5 dose-DLE scenarios ranging from all doses being safe to all doses being very unsafe. We then consider all combinations of these scenarios, resulting in 25 scenarios explored in total. The five dose-DLE and dose-efficacy scenarios for each for ($d_0$, $d_1$, $d_2$, $d_3$) are presented in Table~\ref{tab:single_scenarios}.\\

\begin{table}
\caption{\label{tab:single_scenarios}Safety and efficacy scenarios for ($d_0$, $d_1$, $d_2$, $d_3$).}
\centering
\fbox{\begin{tabular}{ccc}
                    & \textbf{Safety (Probability of DLE, $p_j$)}& \textbf{Efficacy (Hazard Ratio)}  \\ \hline
\textbf{Scenario 0}  & (0.10, 0.12, 0.13, 0.15)  & (1.00, 1.00, 1.00, 1.00)                                                   \\
\textbf{Scenario 1}  & (0.10, 0.12, 0.15, 0.30) & (1.00, 1.00, 1.75, 1.75)                                                    \\
\textbf{Scenario 2}  & (0.10, 0.15, 0.30, 0.45)  & (1.00, 1.50, 1.75, 1.75)                                                   \\
\textbf{Scenario 3}  & (0.10, 0.30, 0.45, 0.60)  & (1.00, 1.50, 1.75, 2.00)                                                   \\
\textbf{Scenario 4}  & (0.10, 0.45, 0.60, 0.60)  & (1.00, 1.75, 2.00, 2.00)                                                   \\ 
\end{tabular}}
\end{table}

We will refer to the scenario with dose-DLE relationship $x$ and dose-efficacy relationship $y$ as ``Scenario $x$--$y$''. Each dose under the combination of DLE and efficacy scenario is classified as incorrect, undesirable, acceptable or desirable. If a treatment is unsafe or has a hazard ratio of 1 then it is classed as incorrect. If it is safe, then a hazard ratio of 1.25 is undesirable, 1.5 is acceptable and at least 1.75 is desirable.\\

For all 25 scenarios, a sensitivity analysis is conducted on varying values of $c_1$ and $c_2$ in order to assess the effect of both altering the allocation ratio between control and active doses, and the total cohort size, $c$. We also study the implications of not sharing controls across doses. A total of six settings are considered, ($c_1=2$ \& $c_2=1$ and $c_1=2$ \& $c_2=2$ for only the settings where controls are shared across doses, with $c_1=4$ \& $c_2=2$ and $c_1=3$ \& $c_2=3$ for the settings both where controls are shared across doses and where they are not). The maximum number of cohorts per dose varies with cohort size in order to maintain the constant maximum total intake per dose level of 72 across that dose level and control.\\

Software in the form of R code used to produce the presented results is available on GitHub (\url{https://github.com/dose-finding/covid19-agile}).

\subsection{{Choice of design parameters}}
\subsubsection{Safety Model}
The proposed design requires the prior and design parameters for both safety and efficacy parts to be pre-specified in advance of the conduct of the trial. The procedure of how these parameters were chosen are given below.\\

The prior parameters for the safety model were obtained via a calibration procedure~\citep{lee2009model} over a number of safety scenarios (not taking into account efficacy). We use safety scenarios 1--3 in Table~\ref{tab:single_scenarios} that correspond to the target dose being $d_3$, $d_2$, and $d_1$, respectively, thus covering various locations of the target dose on the dosage grid.\\

The following prior distribution for the vector of safety model parameters $\theta$ was assumed:
$$(\theta_1, \log(\theta_{2})) \sim \mathcal{N}(\mu, \Sigma)$$
where $\mu= \left (\mu_{1}, \mu_{2} \right)^{\rm T}$ is the vector of means and 
\[
\Sigma=
\left [ \begin{array}{ccc}
\sigma_{1} & \sigma_{12}   \\
\sigma_{12}  & \sigma_{2} \\
\end{array}
\right ].
\]\\

Given the link between the prior toxicity on the control and the intercept parameter $\theta_1$ as  implied by Equation~\ref{eq:skeleton}, $\mu_1= {\rm logit}(\hat{p}_0^{(0)})$ where the prior DLE probability at the control, $\hat{p}_0^{(0)}$. Following discussions with the clinical team, the DLE risk on control was set to  $\hat{p}_0^{(0)}=0.10$. To reduce the computational complexity of the calibration, the covariance between the model parameters was assumed to be $\sigma_{12}=0.$ The rest of the parameters were chosen by conducting simulations using various combinations of values of the parameters on the grid, $\mu_2=\{-0.5,-0.25,0,0.25,0.5\},\sigma_1=\{1.2,1.3,1.4,1.5,1.6 \},\sigma_2=\{0.15,0.25,0.35,0.45,0.55\}.$\\

Furthermore, to define the standardized doses $\tilde{d}_j$ in Equation~\ref{eq:skeleton}, the prior toxicity probability at each dose should be assumed. As there is no reliable information of the DLE rates in the COVID-19 population, the skeleton was also calibrated. The grid of values for the prior toxicity risks are chosen in terms of the difference in the probability of DLE between the neighbouring doses. Specifically, $\hat{p}_j^{(0)}= \hat{p}_0^{(0)} + \nu \times j, j=1,2,3$ where $\nu$ is the difference in the toxicity probabilities between doses, which are then used to find the skeleton using Equation~\ref{eq:skeleton}. The grid of values of $\nu$ was included as one of the parameters for the calibration $\nu=\{0.075,0.100,0.125,0.150\}$. Below, we fix $c_{\rm overdose}=0.25$ that was previously found to result in good safety properties of the design for the two-parameter logistic model~\citep{bailey2009bayesian}.\\

The calibration was performed as follows. For each combination of parameters of $\nu,\mu_2,\sigma_1,\sigma_2$ on the specified of grid, 500 simulations were run under each of the three considered scenarios monitoring the proportion of target dose selections. Then, the selected values of the parameters are those that maximized the geometric mean (taken across scenarios for the same combination of values of the parameters) of the proportion of the target dose selection. This resulted in using  $\nu=0.125,\mu_2=-0.25,\sigma_1=1.40,\sigma_2=0.35$ for the further design evaluation.

\subsubsection{Efficacy Model}

The efficacy stopping boundaries for a particular setting were taken as the pair $(l,u)$ that maximizes 
\begin{equation}\mbox{Criterion} = \mbox{Power} - \lambda \{\mathbb{E}(N_0) + \mathbb{E}(N_1)\}, \label{boundary_crit}\end{equation}
subject to the constraint that the type I error is less than or equal to 10\%, where $\mathbb{E}(N_0)$ and $\mathbb{E}(N_1)$ are the expected sample sizes (across both active dose and control arms) under the null and alternative hypothesis, respectively and $\lambda$ is a tuning parameter. \\

A value of $\lambda = 1/320$ was chosen to allow a power of 80\% to be achieved when $\psi = 1.75$ for the main settings where $n_c=30$ and $c_1:c_2 = 4:2$. For comparability, the same $\lambda$ was used in the sensitivity analysis where the size of cohorts and/or the use of past controls was varied. In these other settings power is lower than 80\%, ranging from 63\% to 79\%. The boundaries for the scenarios are shown in Table \ref{tab:boundaries}.\\

Throughout, the point prior is taken to be $\pi^{(j)}_{E} = \frac{1}{2}$ for all doses $j$. Note that if the boundaries are chosen based on (\ref{boundary_crit}), the operating characteristics of the efficacy design are not affected by the choice of point prior. Specifically, since (\ref{post_prob}) can be re-expressed in terms of the posterior log-odds as
$$\log \left(\frac{\pi^{(j)}_{E|k}}{1 - \pi^{(j)}_{E|k}}\right) = \log \left(\frac{\pi^{(j)}_{E}}{1 - \pi^{(j)}_{E}}\right) + \log\left(\frac{L(\psi^{*}|D^{(j)}_k)}{L(1 |D^{(j)}_k)}\right),$$ 
changing the value of $\pi^{(j)}_{E}$ merely has the effect of translating the posterior log-odds by a constant. Hence the boundaries $(\tilde{l},\tilde{u})$ under the alternative prior will satisfy the relationship $(\mbox{logit}(\tilde{l}),\mbox{logit}(\tilde{u})) = (\mbox{logit}(l) + \xi,\mbox{logit}(u)+\xi)$ where $\xi$ is the log-odds ratio between the new and old prior odds of efficacy. 

\begin{table}
\caption{\label{tab:boundaries}Boundaries for settings in the sensitivity analysis of cohort sizes. The settings have the same maximum sample sizes and common criterion to trade-off power and average sample size.}
\centering
\fbox{
\begin{tabular}{cccccccc}
\multicolumn{2}{c}{Cohort}& & & & & & \\ 
$c_1$ & $c_2$ & $n_c$ & $l$ & $u$ & Power & Criterion \\
\hline
4 & 2 & 30 & 0.224 & 0.839 & 0.800 & 0.630\\
3 & 3 & 30 & 0.268 & 0.841 & 0.747 & 0.573\\
2 & 1 & 30 & 0.192 & 0.858 & 0.794 & 0.634\\
2 & 2 & 30 & 0.227 & 0.858 & 0.744 & 0.566\\
4 & 2 & 0 & 0.317 & 0.815 & 0.634 & 0.438\\
3 & 3 & 0 & 0.271 & 0.821 & 0.691 & 0.484\\
\hline
\end{tabular}
}
\end{table}

\subsection{Results}

\begin{figure}[ht!]
\begin{center}
\includegraphics[width=1\textwidth]{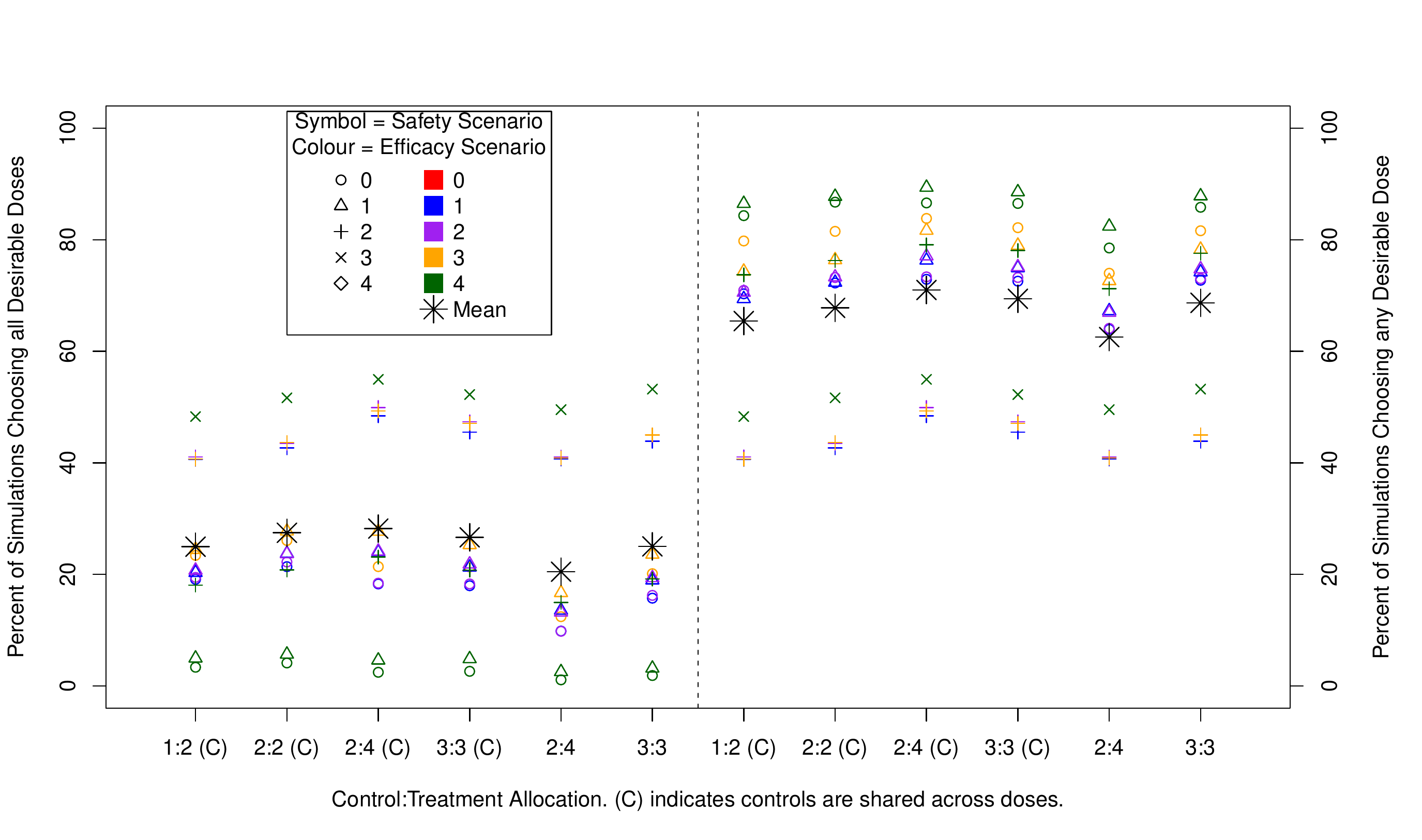}
\end{center}
\caption{Percentage of simulations that recommend all desirable doses (left) and the percentage of simulations that recommend any desirable dose (right) for different cohort size and composition and with and without sharing control group data. Note that only 13 out of 25 efficacy/safety scenarios contain a desirable dose.}
\label{fig:allanycorrect}
\end{figure}

\begin{figure}[ht!]
\begin{center}
\includegraphics[width=1\textwidth]{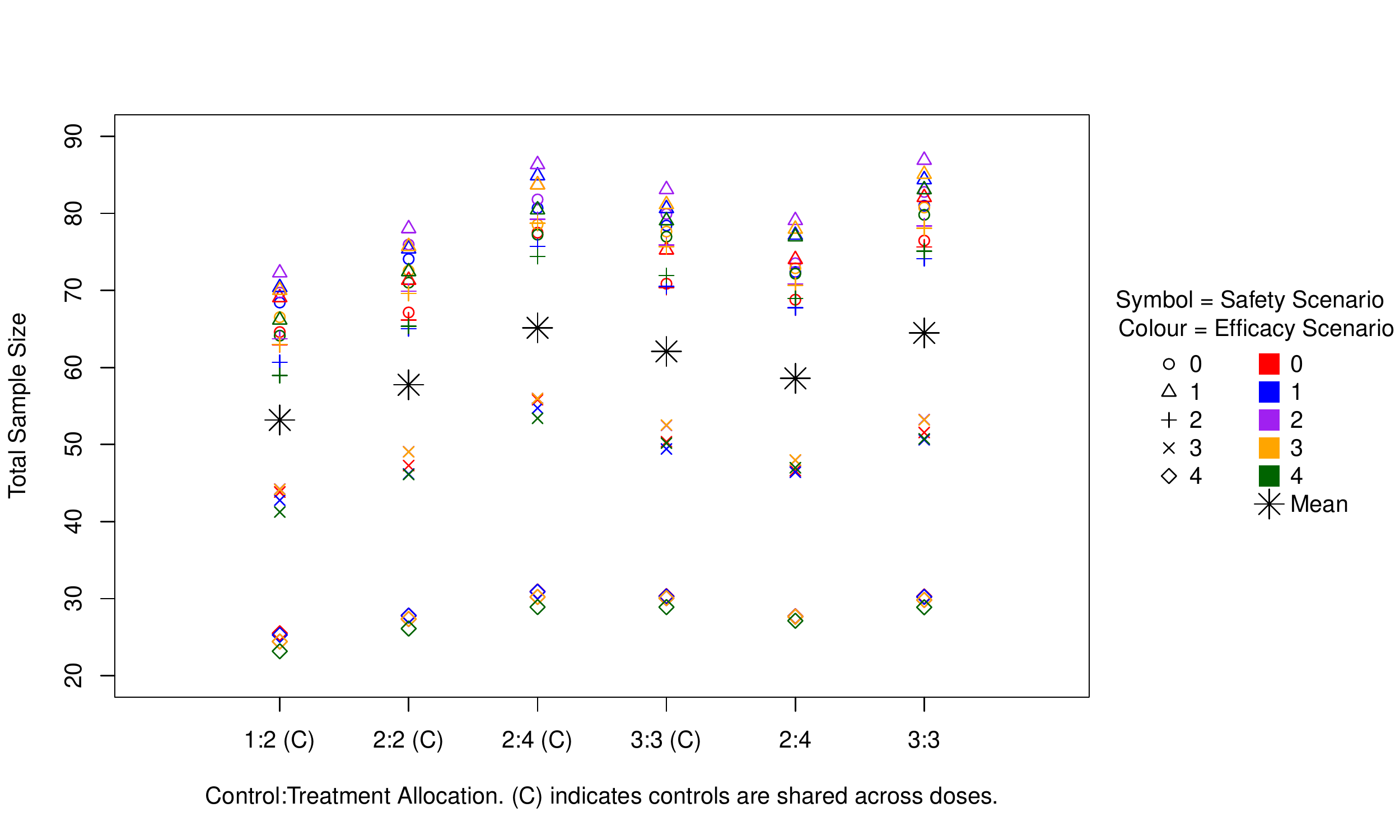}
\end{center}
\caption{Average total sample size across simulations for all scenarios.}
\label{fig:samp_size}
\end{figure}

Detailed results for the setting with cohort size of 4+2 and inclusion of up to thirty control patients are presented in Table \ref{tab:dose_rec}. A comparison of results for the varying cohort sizes are illustrated in Figures \ref{fig:allanycorrect} and \ref{fig:samp_size}. Across settings, the type I error rate, that is the percentage of simulations in Scenario 0--0 where any dose is recommended, ranges from 11\% to 14\%.\\

\begin{table}
\hspace*{-1.5cm}\caption{\label{tab:dose_rec}Percentage of 10,000 simulations where each dose is recommended for ($d_1$, $d_2$, $d_3$) for $c_1=4$ \& $c_2=2$, with  controls shared across doses. Desirable doses are highlighted in \textbf{bold} and Acceptable doses are highlighted in \textit{italics}. Note that these may sum to more than 100\% for each scenario as more than one dose can be recommended simultaneously.}
\centering
\hspace*{-1.5cm}{\footnotesize
\fbox{\begin{tabular}[ht]{ccccccc}
  & \multicolumn{6}{c}{\textbf{Efficacy Scenario}} \\
  \cline{2-7}
  
& & \textbf{0} & \textbf{1} & \textbf{2} & \textbf{3} & \textbf{4} \\ 
  \cline{2-7}
\multirow{2}{*}{\textbf{Safety}} & \textbf{0} & (0.4, 2.3, 8.9) & (0.4, \textbf{24.0}, \textbf{67.2}) & (\textit{2.9}, \textbf{24.2}, \textbf{67.5}) & (\textit{2.9}, \textbf{24.6}, \textbf{80.6}) & (\textbf{5.1}, \textbf{29.7}, \textbf{80.9}) \\ 
\multirow{3}{*}{\textbf{Scenario}}   & \textbf{1} & (1.1, 5.9, 8.0) & (1.2, \textbf{47.4}, \textbf{52.9}) & (\textit{8.6}, \textbf{47.6}, \textbf{53.6}) & (\textit{9.1}, \textbf{47.3}, \textbf{62.1}) & (\textbf{14.0}, \textbf{56.3}, \textbf{61.6}) \\ 
   & \textbf{2} & (5.5, 7.9, 2.4) & (5.7, \textbf{48.4}, 9.4) & (\textit{31.7}, \textbf{49.9}, 9.8) & (\textit{31.4}, \textbf{49.3}, 10.9) & (\textbf{45.0}, \textbf{57.2}, 10.9) \\ 
   & \textbf{3} & (8.4, 2.1, 0.1) & (8.2, 7.3, 0.2) & (\textit{40.4}, 7.8, 0.2) & (\textit{40.4}, 7.9, 0.2) & (\textbf{55.0}, 8.3, 0.2) \\ 
   & \textbf{4} & (5.0, 0.1, 0.0) & (4.9, 0.4, 0.0) & (17.6, 0.3, 0.0) & (17.6, 0.3, 0.0) & (23.1, 0.4, 0.0) \\ 
   \cline{2-7}
\end{tabular}
}}
\end{table}
\normalsize

Figure \ref{fig:allanycorrect} shows the percentage of simulations where all desirable doses are recommended (left) and where any desirable doses are recommended (right). For the baseline setting of $c_1=4$ and $c_2=2$, the mean percentage of simulations recommending all desirable doses is 28.2\%, whilst 71.0\% of simulations recommended any desirable dose.  As expected, safety/efficacy scenarios where only one dose is desirable have higher percentages of simulations recommending all desirable doses, whilst lower percentages of simulations recommending any desirable dose.  The highest power for recommending any desirable dose is observed in the safety/efficacy scenarios where all active doses are safe and efficacious, reaching 89\% for Scenario 1--4.\\

The sensitivity analysis across the varying cohort settings shows there is a very small difference in performance. The ordering of performance across safety/efficacy scenarios is identical with only a small numerical difference. However, we can see that not sharing controls across doses decreases the performance. Power for recommending any desirable dose increases for increasing cohort size $c$, on average increasing 6\% for allocation 1:2 ($c=3$ and $c=6$) and 2\% for allocation ratio 1:1 ($c=4$ and $c=6$). Again unsurprisingly, a lower power is observed when controls are not shared across doses, and within this a higher power when the allocation ratio between control and active dose within cohort is equal. The highest power is achieved for cohorts sizes $c_1=4$ and $c_2=2$ with controls being shared across doses.\\

Figure \ref{fig:samp_size} illustrates the average total sample size across the scenarios and settings. On average about 65 patients are required in the setting used in the trial with the total sample size exceeding 150 in only 1\% of simulations across scenarios. The scenarios with the smallest sample sizes are those where all doses are unsafe and the trial is therefore stopped early for safety. In such cases, it takes 30 patients across 6 weeks on average to reach the conclusion of stopping early for safety. The scenarios with larger sample sizes are those where all doses are safe and most are acceptable or only just desirable (i.e. not the case where the hazard ratio is 2.00), as in these cases more doses are taken to the efficacy part and more patients are required to detect the smaller difference in hazard ratios.\\

It can be seen across settings that the larger the total cohort size, the larger the total sample size. This also corresponds to the higher power settings. When controls are shared across doses, altering the control:treatment allocation from 1:1 to 1:2 decreases the total average sample size, whereas when controls are not shared across doses this increases. This also links to the relationship with the power in the corresponding settings; a higher power is achieved for equal allocations when controls are not shared. The sharing of controls means more patients can be allocated to the novel treatment and hence more can be learnt about it. In all cases, the average total sample size is below 90.\\

Table \ref{tab:dose_rec} gives more detail into which doses are recommended across simulations. For example, in efficacy scenario 2 where the lowest dose is acceptable and the higher two are desirable in terms of efficacy. In safety scenario 0 where all doses are safe, the highest dose is chosen most often. In safety scenario 2 where only the highest dose is unsafe, the middle dose is chosen most often, although less often than the highest desirable dose in scenario 0. In safety scenario 4 where all doses are unsafe, the lowest does is chosen only 17.6\% of the time. It is clear that desirable doses are recommended most often, with incorrect and undesirable doses rarely recommended. This gives insight that the procedure is successful in identifying desirable doses of a single agent.

\section{Extension to combination treatments\label{sec:combomethods}}
\subsection{Setting}

Consider now a randomized controlled dose-escalation dual-agent clinical trial studying the combinations of $J$ doses $d_1<d_2,\ldots,d_J$ of the first compound (referred to as agent $A$) and of $L$ doses $s_1<s_2<\ldots<s_L$ of the second compound (referred to as agent $B$). As before, let $d_0=s_0$ be a zero dose of each compound, respectively, correspondingly to the control treatment, and denote the combination of dose $d_j$ of $A$ and dose $s_l$ of $B$ by $(d_j, s_l)$. Within each agent (assuming the second agent is fixed), it is known prior to the trial that the risk of a DLE monotonically increases with the dose. The objective of the trial is then to study the safety of the combinations and to establish the maximum tolerated combination, the combination corresponding to the target ADLE over the control arm of 20\%. Denoting the probability of DLE at combination with doses $d_j$ and $s_l$  by $p(d_j, s_l)$, the probability of each agent given individually by $p(d_j)$ and $p(s_l)$, and the probability of DLE at the control by $p(d_0, s_0)=p_0$, the aim is to find the combination $(d_{j^\star}, s_{l^\star})$  minimising

$$ | \left(p(d_j, s_l)-p_0 \right) -\gamma|.$$ 

The fundamental difference to the single-agent setting introduced in Section~\ref{sec:single-agent-setting} is that one cannot order all of the combinations of the compounds with respect to the monotonically increasing risk of DLE despite the monotonicity assumption being satisfied within each compound. For example, comparing combination $(d_1,s_2)$ and $(d_2,s_1)$, the dose of one agent is increased and another is decreased, and it is unknown prior to the trial which of these effects prevails in the overall DLE risk associated with the combinations. Consequently, the model-based design for monotherapies in Section~\ref{sec:single} cannot be used and an alternative dose-finding. Below, the extension of the single-agent model-based design is suggested. 

\subsection{Dual-Agent Bayesian Dose-Escalation Model}

For the considered randomized dual-agent combination setting, under the assumption of independence of the compounds, the probability of a DLE associated with combination $(d_j, s_l)$ can be written as
\begin{equation}
p_0(d_j, s_l) = 1 - (1 - p(d_j))(1 -p(s_l)).
\label{ind-combo}
\end{equation}
To allow for the interaction of the compounds in terms of the probability of a DLE, we use the proposed model by~\cite{neuenschwander2015bayesian}
$$odds(p(d_j, s_l)) = odds(p_0(d_j,s_l)) \times \exp \left( \eta \times  \tilde{d}_j \times 
\tilde{s}_l \right),$$
where  $odds(p)=\frac{p}{1-p}$ is the odds transformation of the probability $p$, and ${\eta}$ is the interaction coefficient, positive values of which correspond to synergistic DLE risk, zero corresponds to additive effect without interaction, and negative values correspond to antagonistic risk of DLE, and  $\tilde{d}_j$, $\tilde{s}_l$ are standardized dose levels corresponding to the same DLE probability as the doses ${d}_j,s_l$.\\

Note that $p(\cdot)$ in Equation~(\ref{ind-combo}) is the probability of a DLE associated with one compound given as monotherapy as in Section~\ref{sec:single}. Therefore, we adopt the 2-parameter logistic model given in Equation~(\ref{eq:toxicity}) for each agent separately. Specifically, let 
\begin{equation}
p(d_j)=\psi(\tilde{d}_j,\theta_{1},\theta_{21})
\label{eq:combo1}
\end{equation}
and 
\begin{equation} p(s_l)=\psi(\tilde{s}_l,\theta_{1},\theta_{22})
\label{eq:combo2}
\end{equation}
where $\theta=(\theta_1,\log \left( \theta_{21}\right), \log \left( \theta_{22} \right),\eta)$ are the unknown parameters with a Normal prior distribution,  $\theta \sim \mathcal{N}(\mu, \Sigma)$
where $\mu= \left (\mu_{1}, \mu_{21}, \mu_{22}, \mu_\eta \right)^{\rm T}$ is the vector of means and 
\[
\Sigma=
\left [ \begin{array}{cccc}
\sigma_{1} & \sigma_{1,21}  &  \sigma_{1,22} & 0 \\
\sigma_{1,21}  & \sigma_{21} &  0 & 0 \\
\sigma_{1 ,22} &  0 & \sigma_{22} & 0 \\
0 & 0 & 0& \sigma_\eta\\
\end{array}
\right ]
\]
As before, we require the standardized dose level corresponding to the control treatment to be equal to $\tilde{d}_0=\tilde{s}_0=0$, so that the intercept parameter of the 2-parameter model~(\ref{ind-combo}) relates to the probability of DLE on the control only. Therefore, both single-agent models employ the same intercept parameter $\theta_1$ as it corresponds to the probability of a DLE at the same control treatment. Consequently, for small to moderate values of probability of DLE,  the intercept parameter in the single-agent model, $\theta_1$ approximately equals the logit inverse-logit transformation of the half of the probability of DLE on the control treatment subject ${\rm logit} \left( \frac{p(d_0,s_0)}{2} \right) = {\theta_1}$. This is used to construct the standardized dose levels $\tilde{d}_j$, $\tilde{s}_l$ using the prior means of the parameters $\theta$ and the prior probabilities of a DLE at each combination similarly to the construction in Equation~(\ref{eq:skeleton}).\\

Parameters of the vector $\theta$ are the unknown quantities that define the combination-DLE relationship. As in the single-agent design in Section~\ref{sec:single}, the posterior distribution of these are sequentially updated using the data collected during the trial using Bayes' theorem. Specifically, denote the joint prior distributions of vector $\theta$ by $f_0(.)$. Assume that $n$ patients have received the combinations $(d1),s(1)) \ldots, (d(n),s(n))$ and binary responses $\mathbb{Y}_n=[y_1, \ldots, y_n]^{\rm T}$ were observed, respectively. The models update the posterior distribution of $\theta$ as
\begin{equation}
f_n(\theta)= \frac{f_{n-1}(\theta)\phi((d(n),s(n)),y_n,\theta)}{\int_{\mathbb{R}^4} f_{n-1}(u)\phi((d(n),s(n),y_n,u) {\rm d} u}=\frac{f_0(\theta) \prod_{i=1}^{n}\phi((d(i),s(i),y_i,\theta)}{\int_{\mathbb{R}^4}  f_0(u) \prod_{i=1}^{n}\phi((d(i),s(i),y_i,u) {\rm d} u}
\label{posterior2}
\end{equation}
where
$$\phi((d(n),s(n),y_n,\theta)=p(d(n), s(n), \theta)^{y_n}(1-p(d(n), s(n), \theta))^{1-y_n}.$$
This posterior distribution is then used to make the escalation and de-escalation decision during the trials as proposed below.

\subsection{Dual-Agent Dose-Escalation Design}

The above combination-DLE model is then used in the design in Section~\ref{design} in place of the single-agent model. As the efficacy part of the dose-escalation design proposed for monotherapies considered each dose individually, the efficacy part of the combination study remains the same. Once the combination of the compounds is established to be safe, it  is graduated into the efficacy part following the single-agent proposal and the same decision rules for dropping for futility and safety. 

\section{Evaluation of combination treatment design \label{sec:comboresults}}
\subsection{Scenarios}

In order to evaluate the dual-agent design, we conduct a simulation study comprising of scenarios with two dose levels of agent $A$ ($d_1$ \& $d_2$) and three dose levels of agent $B$ ($s_1$, $s_2$ \& $s_3$). We consider four dose-DLE scenarios ranging from a situation where all combinations are safe to a case where all are unsafe. Four dose-efficacy scenarios ranging from no efficacious combination to a steep monotonic within agent relationship are considered yielding 16 safety-efficacy scenarios. The dual agent scenarios are presented in Table~\ref{tab:combo_scenarios} with definitions of incorrect, undesirable, acceptable and desirable dose combinations remaining as they were previously defined for single agent doses in Section \ref{sec:scen_single}. Here we fix cohort sizes to the previous baseline setting of $c_1=4$ (assigned to active dose combinations) and $c_2=2$ (assigned to the control) and allow controls to be shared across dose combinations. The maximum number of patients per dose combination is 72 as before.

\begin{table}
\caption{\label{tab:combo_scenarios}Safety and efficacy scenarios for dual-agent combinations $A$ and $B$. It is assumed the control arm remains with probability of DLE 0.10 and hazard ratio 1.00.}
\fbox{
\begin{tabular}{ccrlrl}
\hline
                                     & \multicolumn{3}{c}{\textbf{Safety}}                       & \multicolumn{2}{c}{\textbf{Efficacy}}       \\
\multicolumn{1}{l}{}                 & \multicolumn{3}{l}{\textbf{(Probability of DLE, $p_j $)}} & \multicolumn{2}{c}{\textbf{(Hazard Ratio)}} \\ \hline
\multicolumn{1}{l}{}                 &                    & $d_1$           & $d_2$          & $d_1$              & $d_2$              \\
\multirow{3}{*}{\textbf{Scenario 0}} & $s_1$            & \hspace{10pt}0.10              & 0.12             & \hspace{10pt}1.00                 & 1.00                 \\
                                     & $s_2$            & 0.13              & 0.15             & 1.00                 & 1.00                 \\
                                     & $s_3$            & 0.15              & 0.18             & 1.00                 & 1.00                 \\ \hline
\multirow{3}{*}{\textbf{Scenario 1}} & $s_1$            & 0.10              & 0.12             & 1.00                 & 1.25                 \\
                                     & $s_2$            & 0.25              & 0.30             & 1.25                 & 1.50                 \\
                                     & $s_3$            & 0.50              & 0.55             & 1.50                 & 1.75                 \\ \hline
\multirow{3}{*}{\textbf{Scenario 2}} & $s_1$            & 0.15              & 0.30             & 1.00                 & 1.50                 \\
                                     & $s_2$            & 0.25              & 0.35             & 1.25                 & 1.75                 \\
                                     & $s_3$            & 0.30              & 0.45             & 1.50                 & 2.00                 \\ \hline
\multirow{3}{*}{\textbf{Scenario 3}} & $s_1$            & 0.40              & 0.45             & 1.00                 & 1.50                 \\
                                     & $s_2$            & 0.45              & 0.50             & 1.50                 & 1.75                 \\
                                     & $s_3$            & 0.50              & 0.55             & 1.75                 & 1.75                 \\ \hline
\end{tabular}
}
\end{table}

\subsection{Parameter Calibration}
To define the parameters of the combination model, a calibration procedure similar to the procedure described in Section~3.3.1 was applied. Safety scenarios 0--3 in Table~\ref{tab:combo_scenarios} that correspond to different steepness of the combination-toxicity relationship and different locations of the target combination were used. We then choose the hyperparameters for the prior distribution of the parameters of the model, $\theta=(\theta_1,\log \left( \theta_{21}\right), \log \left( \theta_{22} \right),\eta)$, given in Equation~(\ref{eq:combo1}) and Equation~(\ref{eq:combo2}).\\

Given the link between the prior toxicity on the control arm, and the intercept parameter $\theta_1$ that is common for both single-agent parameter models, we set $\mu_1= {\rm logit}\left(\hat{p}_0^{(0)})/2\right)$ where $\hat{p}_0^{(0)}=0.10$ is the DLE risk on the control, as before.  To reduce the computational complexity, the covariance between all model parameters was assumed to be $\sigma_{1,21}=\sigma_{1,22} =0$, and the variance of the slope parameters in each single-agent model is the same $\sigma_{21}=\sigma_{22}.$ Finally, we set the mean of the distribution of the interaction parameter $\mu_{\eta}=0$ to reflect that either synergistic and antogonistic effects are possible.   The rest of the parameters were chosen by simulations using combinations of values on the grid, $\mu_{21}=\{-0.4,0.2,0.0,0.2 \}, \mu_{22}=\{-0.4,0.2,0.0,0.2  \}, \sigma_{1}=\{0.4,0.6,0.8,1.0,1.2 \}, \sigma_{21}=\sigma_{22}=\{0.05,0.15,0.25,0.35,0.45\}$ and $\sigma_{\eta}=\{0.02,0.04,0.10,0.20,1\}.$\\

As in the single-agent setting, the standardized doses $\tilde{d}_j, \tilde{s}_l$ were calibrated in terms of the difference in the probability of DLE between the neighbouring doses. Specifically, $\hat{p}_j^{(0)}(d_j)= \hat{p}_0^{(0)} + \nu_{d} \times j,$ and $\hat{p}_j^{(0)}(s_l)= \hat{p}_0^{(0)} + \nu_{s} \times j$, $l,j=1,2,3$ where $\nu_{d}, \nu_{s}$ are the differences in the toxicity probabilities between doses of the first and second agents, respectively. The following values of differences were tried $\nu_{d}=\{0.025, 0.05,0.075,0.10,0.125\}$, $\nu_{s}=\{0.025, 0.05,0.075,0.10,0.125\}$.  Finally, we fix $c_{\rm overdose}=0.25$ for the overdose control constraint. \\

Using 500 simulations under each scenario and each combination of hyperparameters, the values $\mu_{21}=0.0 , \mu_{22}=0.0, \sigma_{1}=0.6, \sigma_{21}=\sigma_{22}=0.25$, $\sigma_{\eta}=0.10, \nu_{d}=\nu_{s}=0.075$ were found to maximize the geometric mean of the proportion of correct combination selections across the scenarios.

\subsection{Results}

\begin{figure}[ht!]
\begin{center}
\includegraphics[width=1\textwidth]{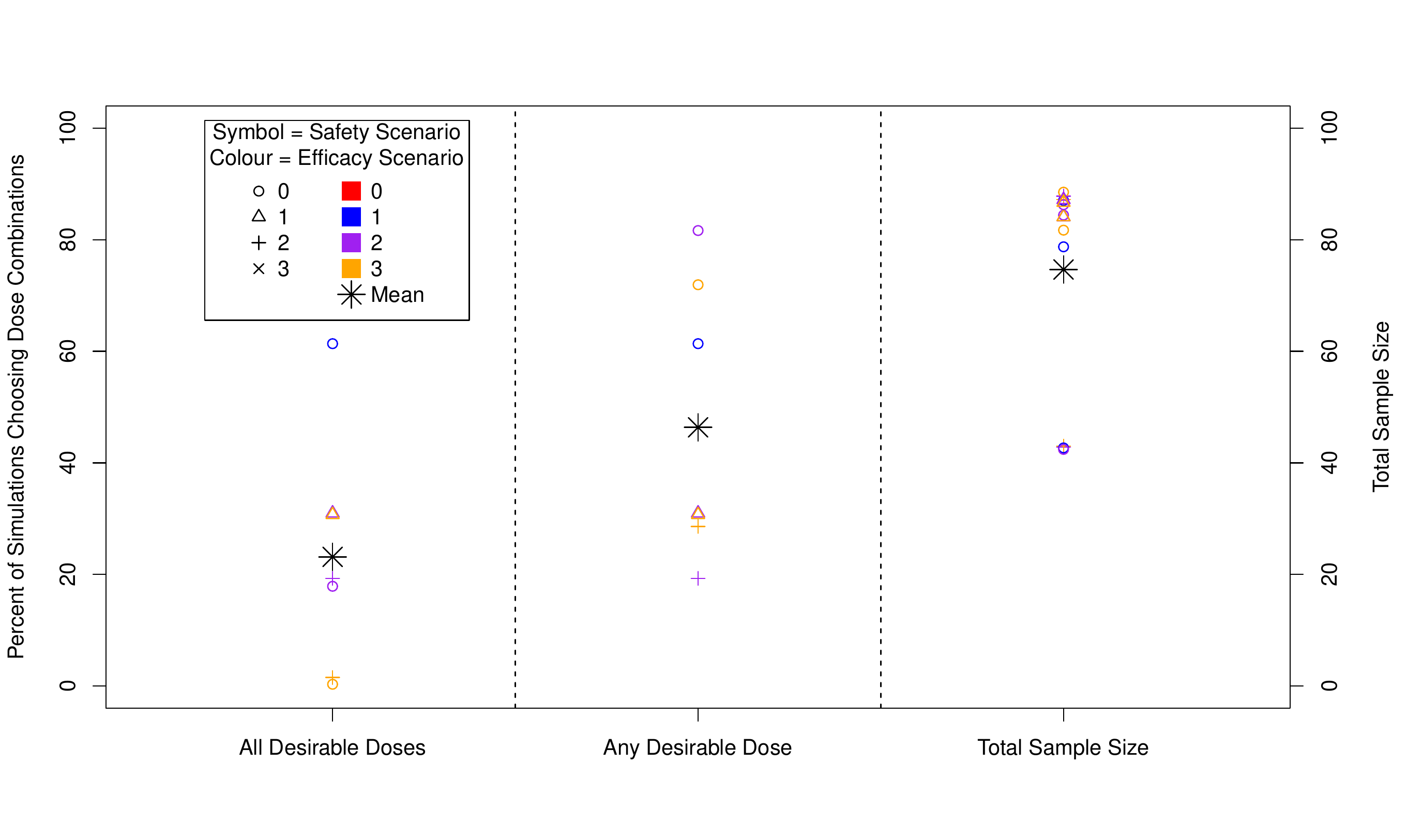}
\end{center}
\caption{Percentage of 10,000 simulations that recommend all desirable dose combinations (left), the percentage of simulations that recommend any desirable dose combination (centre) and average total sample size. Note that only 7 out of 16 efficacy/safety scenarios contain a desirable dose combination.}
\label{fig:combo_res}
\end{figure}

For the 16 scenarios considered in the simulation study, the percentage of simulations recommending any desirable dose combination, the percent of simulations recommending all correct dose combinations and the mean total sample size are presented in Figure~\ref{fig:combo_res}, with further detail on individual dose combination recommendations given in Table~\ref{tab:combo_rec}. The type I error rate, that is the percentage of simulations recommending any dose combination in Scenario 0-0, is 12.1\%. In the extension from monotherapies to dual agent therapies, some similar patterns are maintained in the results although there are some notable differences. \\

It can be seen in Figure~\ref{fig:combo_res} that the spread of powers across scenarios for dose combinations is larger than for monotherapies, for both the selection of all and any desirable dose combinations. The maximum power of 81.7\% to select any desirable dose combination is achieved in Scenario 0-2, where all doses are safe and there is a steep monotonic relationship within agent for efficacy. Even though there is an extra desirable dose combination in Scenario 0-3, we observe a slightly reduced power since the most efficacious dose combination has a lower hazard ratio. The power to recommend all desirable doses ranges up to 61\% in Scenario 0-1 where only one dose is desirable, with the lowest in Scenarios 0-3 and 2-3 where the desirable dose combinations are across separate agents' dose escalation (i.e. to select all desirable doses requires a de-escalation in one agent and then escalation in the other).\\

Across scenarios, the mean total sample size is 75, ranging from 42-89, a narrower range than for the single agent. However similar to the single agent, the smaller sample sizes correspond to scenarios where all dose combinations are unsafe and therefore the trial stopped early for safety. When this is not the case, there is little variation across scenarios in terms of mean total sample size.\\

Table~\ref{tab:combo_rec} shows further details of dose recommendations in the simulations. Especially of note is the emphasis on recommendations of acceptable doses. For example in Scenario 2-3 where the power to detect desirable dose combinations is low, a large proportion of simulations also recommend an acceptable dose combination. It is also clear that inefficacious and/or unsafe doses are rarely recommended.\\

\begin{table}
\caption{\label{tab:combo_rec}Percentage of 10,000 simulations where each dose combination is recommended. Desirable dose combinations are highlighted in \textbf{bold} and Acceptable dose combinations are highlighted in \textit{italics}. Note that these may sum to more than 100\% for each scenario as more than one dose combination can be recommended simultaneously.}
\centering

\fbox{\begin{tabular}{cccccccccc}
                                              &                             & \multicolumn{8}{c}{\textbf{Efficacy Scenario}}                                                                                   \\ \cline{2-10} 
                                              &                             & \multicolumn{2}{c}{\textbf{0}} & \multicolumn{2}{c}{\textbf{1}} & \multicolumn{2}{c}{\textbf{2}} & \multicolumn{2}{c}{\textbf{3}} \\ \cline{2-10} 
\multirow{5}{*}{} & \multirow{3}{*}{\textbf{0}}  & 0.2 & 0.8 & 0.2 & 2.5 & 0.1 & \textit{5.1} & 0.2 & \textit{5.0} \\ 
  & & 0.8 & 2.9 & 2.7 & \textit{17.3} & 2.5 & \textbf{25.2} & \textit{5.0} & \textbf{25.1} \\ 
  & & 0.3 & 7.9 & \textit{1.6} & \textbf{61.4} & \textit{1.4} & \textbf{74.4} & \textbf{2.2} & \textbf{61.3} \\  \cline{2-10} 
                                              & \multirow{3}{*}{\textbf{1}}   & 1.1 & 3.0 & 1.2 & 9.5 & 1.2 & \textit{17.3} & 1.3 & \textit{16.6} \\ 
  & & 5.5 & 5.3 & 15.2 & \textit{23.3} & 15.1 & \textbf{31.0} & \textit{28.1} & \textbf{30.7} \\ 
 \textbf{Safety} &  & 0.5 & 0.5 & 1.0 & 1.7 & 1.2 & 1.9 & 1.5 & 1.7 \\  \cline{2-10} 
                     \textbf{Scenario}                         & \multirow{3}{*}{\textbf{2}}   & 2.8 & 3.2 & 3.2 & 9.3 & 3.1 & \textit{15.8} & 3.1 & \textit{15.8} \\ \multirow{5}{*}{}
  & & 5.7 & 4.0 & 14.7 & \textit{14.8} & 15.0 & \textbf{19.3} & \textit{26.2} & \textbf{19.0} \\ 
  & & 2.1 & 0.7 & \textit{8.6} & 3.6 & \textit{8.2} & 4.1 & \textbf{11.1} & 3.6 \\   \cline{2-10} 
                                              & \multirow{3}{*}{\textbf{3}}     & 6.7 & 1.1 & 6.5 & 2.2 & 6.6 & 2.8 & 6.6 & 2.9 \\ 
  & & 1.0 & 0.1 & 2.1 & 0.2 & 2.1 & 0.2 & 2.9 & 0.2 \\ 
  & & 0.0 & 0.0 & 0.1 & 0.0 & 0.1 & 0.0 & 0.1 & 0.0 \\
  \cline{2-10}              
\end{tabular}
}

\end{table}

\section{Discussion\label{sec:discussion}}

We introduce and evaluate the statistical design of the AGILE platform which seeks to quickly establish safe doses and potential for efficacy. We find that the design can identify potential treatments with good accuracy and show that the approach is easily extended to combinations of treatments.\\

The design uses a recently proposed randomized dose-finding design to ensure that differences between symptoms of COVID can be distinguished from side-effects of the investigated treatment while a very simple Bayesian model is used to capture the potential efficacy of the treatments. The latter is in line with the objective of the trial: make reliable decisions about potential quickly, rather than using more complex methods that allow more precise estimation. At the same time this approach guaranteed that the whole platform structure could be simulated quickly to enable the study design to be fixed quickly.\\

The design has been constructed in a flexible manner using a time-to-event outcome and we based our simulations on time-to-improvement $-$ an endpoint that has been shown recently to be a highly powered and relatively easy to collect endpoint \citep{dodd2020}. The platform has, however, been constructed to also be able to investigate mild disease in which case a primary endpoint used would be time-to-negative viral titres in nose and/or throat swab. Provided that the event rate in this setting is the same, we expect that the performance reported here will be similar. \\

\section*{Acknowledgement}
T Jaki received funding from UK Medical Research Council (MC\_UU\_0002/14). “This report is independent research supported by the National Institute for Health Research (NIHR Advanced Fellowship, Dr Pavel Mozgunov,
NIHR300576; and Prof Jaki’s Senior Research Fellowship, NIHR-SRF-2015-08-001). The views expressed in this publication are those of the
authors and not necessarily those of the NHS, the National Institute for
Health Research or the Department of Health and Social Care (DHCS).

\bibliographystyle{rss}
\bibliography{COVIDbib}

\end{document}